\documentclass[epjST]{svjour}
\usepackage{graphics}
\usepackage{bm}
\usepackage{amsmath}
\usepackage[usenames, dvipsnames]{color}
\begin{document}


\title{Two-dimensional colloidal mixtures in magnetic and gravitational fields}

\author{H.~L\"owen \and T.~Horn \and T.~Neuhaus \and B.~ten~Hagen}

\institute{Institut f\"ur Theoretische Physik II: Weiche Materie, 
Heinrich-Heine-Universit\"at D\"usseldorf, 
40225 D\"usseldorf, Germany}

\abstract{ 
This mini-review is concerned with two-dimensional colloidal mixtures exposed to various 
kinds of external fields. By a magnetic field perpendicular to the plane,  
dipole moments are induced in paramagnetic particles which give rise to repulsive interactions leading to complex crystalline alloys in the 
composition-asymmetry diagram. A quench in the magnetic field
induces complex crystal nucleation scenarios.
 If exposed to a gravitational field, these mixtures exhibit a brazil--nut effect
and show a boundary layering which is explained in terms of a
 depletion bubble picture. The latter persists for time-dependent gravity (``colloidal shaking''). 
Finally, we summarize crystallization effects when the second species is 
frozen in a disordered matrix which provides obstacles for the crystallizing 
component.
}

\maketitle

\section{Introduction}
\label{sec:intro}


When colloids are confined to interfaces 
\cite{ivlev2012complex,Lowen2001,Lowen2009}, almost perfect two-dimensional		
systems can be realized \cite{Maret_PRL_1997,Zahn1999,PalbergJPCM2005}. The key idea is to consider a pending water
droplet at a glass plate which is filled with superparamagnetic particles, 
see also \cite{EPJEST_15}. Due to gravity
the particles sediment down until they meet the air-water interface. Since the surface tension
of the air-water interface is high, the particles do not penetrate through the interface as
this would create additional interfacial area. Consequently, the particles are confined
to the interface by a combination of gravity acting downwards and interfacial free energy
keeping them upwards. Since the water droplet containing the colloidal particles is
macroscopic, its air-water interface is flat on a micron scale typical for an interparticle
spacing between the colloids.

\begin{figure}
\centering \resizebox{0.5\columnwidth}{!}{
\includegraphics{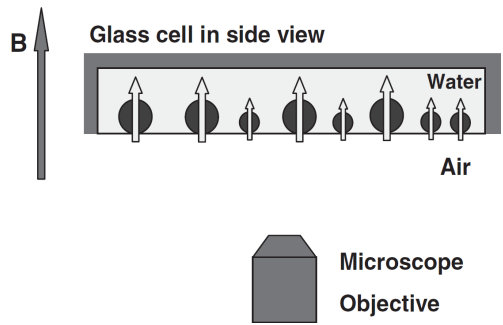} }
\caption{Schematic view of the set-up: binary mixture of superparamagnetic
colloidal particles at an air-water interface in an
external magnetic field $B$ perpendicular to the plane. From Ref. \cite{Hoffmann_PRL_2006}.}
\label{fig:1}
\end{figure}

This set-up can be combined with an external magnetic field $B$, which induces magnetic dipole moments in the particles, $m \propto B$. This in turn results in a dipole-dipole
interaction between the particles. All dipole moments are along the magnetic
field direction. If the external magnetic field is perpendicular to the interface, there
is a repulsive interaction between the particles which can be described by a pairwise
potential scaling as the inverse cube of the distance \cite{Froltsov1}. The prefactor scales with
the square of the magnetic dipole moment, i.e., with the square of the magnetic field
strength $B$. By tuning the magnetic field $B$, one can thus readily change the interaction
strength. For inverse power-law interactions, this corresponds formally to a change of
temperature or density. The experimental set-up is typically combined with video-microscopy in
order to visualize the individual particle trajectories. A schematic view for the set-up in case of
 a binary suspension is shown in Figure \ref{fig:1}.

In the last decades, progress has been achieved for one-component and two-component systems
which is outlined as follows: for one-component systems, the Kosterlitz--Thouless--Halperin--Nelson--Young scenario was confirmed on this strictly two-dimensional
system \cite{Maret_PRL_1999}. Moreover, it was verified that Young's modulus approaches
$16\pi$ at the melting temperature \cite{Gruenberg2004} as predicted by the theory. The modulus of
orientational stiffness was measured in the hexatic phase at the fluid-hexatic transition
and found to be in agreement with theory, too \cite{Keim2007}. Since the creation of disclinations
and dislocations is crucial for the melting scenario, the pair interaction of dislocations
has been determined in these two-dimensional crystals \cite{Eisenmann2005}. Finally, the crystal phonon
dispersion relations \cite{Keim04} have been determined and found to be in full quantitative agreement
with theoretical calculations \cite{Eisenmann2004}. For the crystallization behavior \cite{Loewen_1994}, dynamical density functional theory
was developed and applied to magnetic particles \cite{seed,Sven}.

If the external magnetic field is tilted relative to the surface normal, anisotropic dipole-dipole
interactions between the particles result. The zero-temperature phase diagram
was calculated by lattice sums \cite{Froltsov1} revealing a wealth of anisotropic stable solid lattices in
agreement with experimental data. The Lindemann parameters in the anisotropic
crystals were determined in good agreement between experiment and theory \cite{Froltsov2}. The
melting of the anisotropic crystals is again mediated by defects \cite{Eisenmann_Maret} as in the isotropic
case and the resulting intermediate phase can be called ``smectic-like''.

Two-component (binary) systems with big and small magnetic dipole moments
represent ideal glass formers in two spatial dimensions \cite{konig2005}. Several structural and
dynamical features of these mixtures have been explored including the long-time self
diffusion \cite{Naegele_EPL_2002} and the partial clustering of the small particles at moderate interaction
strengths \cite{Hoffmann_PRL_2006}. The latter is revealed by an unusual peak in the partial structure factor of
the small-small pair correlations. 

In this paper, we review more recent progress obtained by theory and simulation for
two-dimensional colloidal binary mixtures in magnetic and gravitational fields
and compare the results to experimental studies \cite{EPJEST_15}.
The mini-review is organized as follows: in sec~\ref{sec:crystallization} we  start with bulk phase behavior of binary
dipolar mixtures. Real-space experimental data for the partial pair 
correlations are compared to simulations for a binary mixture and good agreement is found.
Moreover the equilibrium phase diagram is discussed and an ultrafast quench to cool down the system 
is described. Then effects of confinement are presented briefly, too. In sec.~\ref{sec:2dmix}
the combination of a magnetic field and perpendicular in-plane gravity is studied and 
a colloidal brazil--nut effect is found and explained in terms of a simple
 effective Archimedian theory. Finally, in sec.~\ref{sec:quencheddisorder}, we consider binary mixtures of 
mobile and immobile particles and explore the freezing transition in this system with 
quenched disorder. We conclude in sec.~\ref{sec:conclusions}.

\section{Crystallization of two-dimensional colloidal mixtures in magnetic fields}
\label{sec:crystallization}

A binary mixture of super-paramagnetic colloidal particles pending at an air-water interface 
is  an excellent realization of a 2D classical many-body system 
\cite{konig2005,Naegele_EPL_2002,Hoffmann_PRL_2006,Hoffmann_JPCM_2006}.
An external magnetic field $B$ perpendicular to the interface induces parallel dipole 
moments $m_{\rm A}$ and $m_{\rm B}$ in particles A and B, respectively, resulting in
an effective repulsive interaction which scales as the inverse cube of the distance 
$r$ within the monolayer. By defining the magnetic susceptibilities per 
particle A and B as $\chi_{\rm A,B} = m_{\rm A,B}/B$, we obtain the pair potential,
\begin{equation}
\label{assoud_eq_V_dip}
V_{\alpha\beta}(r)=\chi_{\alpha}\chi_{\beta}\frac{B^2}{r^{3}}.
\end{equation}
Note that only for low $B$ the induced dipole moment is linearly proportional to the 
external field, and then $\chi_{\alpha}$
is field-independent. In this case, for a fixed relative composition $x_{\rm B}$ and 
susceptibility ratio $\tilde m\equiv m_{\rm
B}/m_{\rm A}=\chi_{\rm B}/\chi_{\rm A}$, all static quantities depend solely on 
the coupling parameter \cite{hansen},
\begin{equation}\label{assoud_eq4}
\Gamma=\frac{\chi_{\rm A}^2B^2}{k_{\rm B}T\Delta_{\rm A}^3},
\end{equation}
where $k_{\rm B}T$ is the thermal energy and $\Delta_{\rm A}=(n_{\rm A})^{-1/2}$ is the mean distance between particles A
\cite{Lowen2012_Advances}, with $n_{\rm A}$ denoting the partial number density of A-particles.

\subsection{Fluid pair structure}

Structural correlations of binary mixtures have been studied in great detail \cite{Naegele_EPL_2002,Assoud_JPCM,assoud2009,Ebert_EPJE_2008}. Figure 2 shows a recent 
comparison between experimental and simulation data for the three partial pair correlation
functions $g_{\rm AA}(r)$, $g_{\rm AB}(r)$, and $g_{\rm BB}(r)$  \cite{Assoud_JPCM}.
Except for a fine substructure in $g_{\rm BB}(r)$, there is very good overall agreement.
 Higher-order structural
correlations, e.g., particles with a square-like and triangular-like surrounding have also been studied
\cite{Assoud_JPCM,Koenig_EPL_2005}. These building blocks are essential for 
understanding the onset of glass formation.

\begin{figure}
\centering \resizebox{0.8\columnwidth}{!}{
\includegraphics{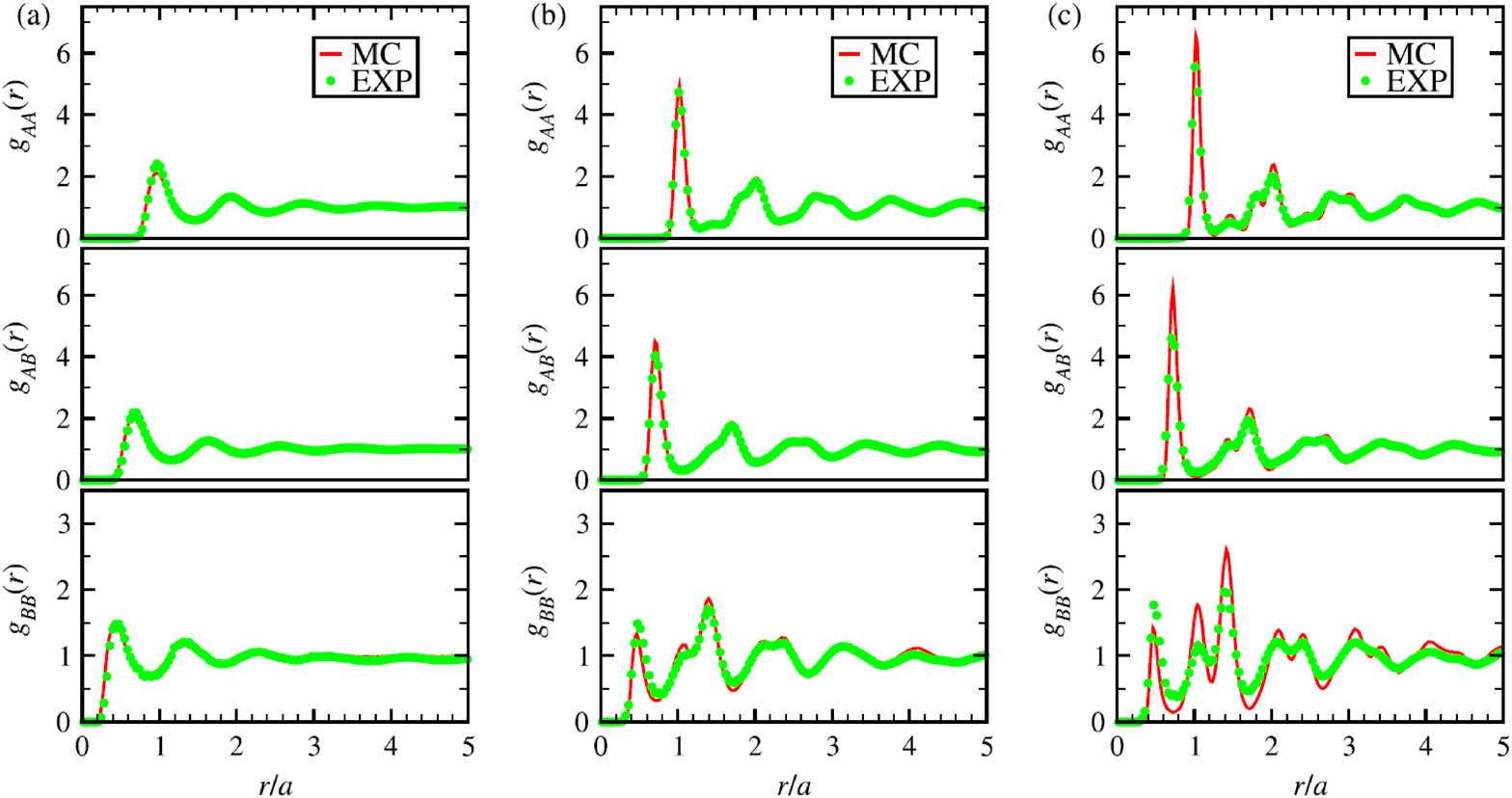} }
\caption{Partial radial pair distribution functions $g_{AA}(r )$, $g_{BB}(r )$, and $g_{AB}(r )$. 
Experimental data (EXP) are compared to Monte Carlo simulation results
(MC) for three different dimensionless coupling strengths $\Gamma$. 
(a) $\Gamma = 4.9$, (b) $\Gamma = 38.9$, and (c) $\Gamma = 82.9$. 
The relative composition of B--particles  is fixed at
$x_{\rm B}  = 0.29$. From Ref. \cite{Assoud_JPCM}.}
\label{fig:2}
\end{figure}

\subsection{Equilibrium bulk phase diagram}

At zero temperature (i.e., for $\Gamma\to\infty$), the state of the binary system is completely determined by the susceptibility ratio
$\tilde m$ (varying in the range $0\leq \tilde m\leq1$) and the relative composition $x_{\rm B}$ of species $B$ (with smaller
dipole moment). A wealth of different
stable phases occur. The topology of the phase diagram is getting more complex with increasing asymmetry \cite{Assoud_2007}, see also  \cite{Fornleitner_SM_2008,Fornleitner2009_Langmuir}. 
For small asymmetries $\tilde m$ and intermediate compositions $x_{\rm B}$, 
the system splits into triangular A$_2$B and
AB$_2$ phases. Experiments with colloidal dipolar mixtures, which were performed for a strong asymmetry of $\tilde m\simeq0.1$, confirmed
the predicted crystalline structures \cite{Ebert_EPJE_2008}, however, only in the form of small crystalline patches. 
Recent results were also obtained for the phonon band structure \cite{Fornleitner2010_PRE} and crystal structures in tilted magnetic fields \cite{Chremos2009_JPhysChemB}. For another application to colloid polymer mixtures see \cite{Lonetti2011_PRL}.

\subsection{Ultrafast quenching}

By suddenly increasing the magnetic field, the system
can be quenched on a time scale which is much smaller than single particle motion
\cite{Dillmann2008}. Since magnetic field strength corresponds formally to an inverse temperature, an ultrafast
temperature quench can be realized experimentally, which is very difficult for molecular
systems.
Analyzing the particle configuration after a rapid quench reveals some local
crystalline patches in the glass \cite{Ebert_EPJE_2008,Assoud_2007}. These patches correspond to the
thermodynamic bulk crystal \cite{chowdhury1985laser}, demonstrating an interplay between vitrification and
crystallization
\cite{Sperl_PRE_2007,Onuki_PRE_2007,Harrowell_PRL_2006,Tanaka_PRL_2007}. 
Experimental snapshots just after the quench and well after 
the quench are shown in Figure 3. 
Within the allotted time, the binary mixture does not find its true ground state
but shows patches with local square and triangular order. The fraction of particles 
with this local order grows with time, see Figure 3, and there is good agreement between
Brownian dynamics simulations and real-space experiments \cite{assoud2009}.

\begin{figure}
\centering \resizebox{0.8\columnwidth}{!}{
\includegraphics{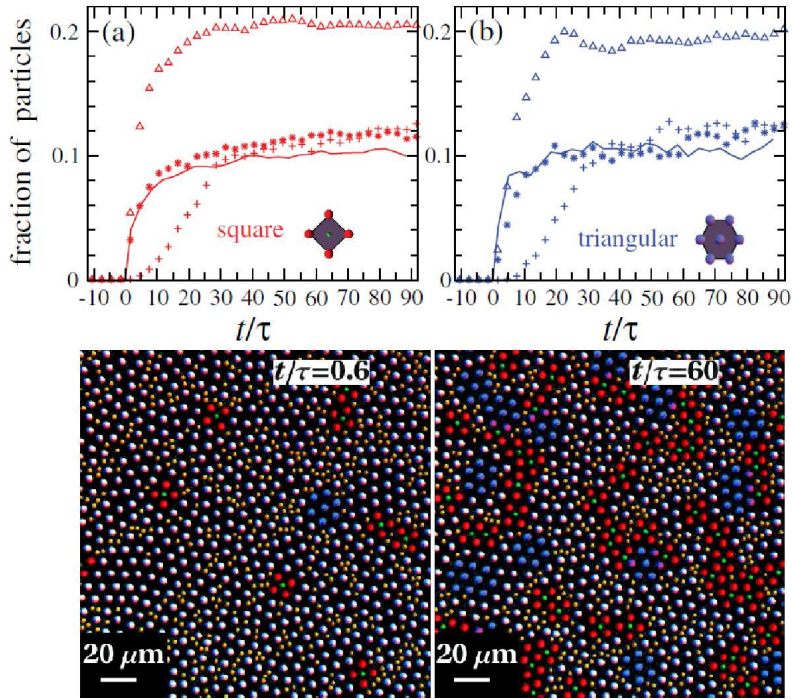} }
\caption{(a) Fraction of B-particles belonging to a
crystalline square surrounding (see inset) and (b) fraction of A-particles belonging to a crystalline triangular surrounding (see
inset) versus reduced time after an ultrafast quench. The lines are experimental data while the symbols
(*) are data from Brownian dynamics simulations. Two experimental snapshots
for a time  just after the quench (left configuration) and
a later time (right configuration) are shown. Big
particles are shown in blue if they belong to a triangular
surrounding and in red if they belong to a square surrounding.
All other big particles are shown in white. Few big particles
belonging to both triangular and square surroundings are shown
in pink. The small particles are shown in green if they belong to a
square center of big particles, otherwise they appear in yellow.
Also included are simulation data for an instantaneous ``steepest
descent'' quench ($\triangle$) and for a linear
increase of the field  ($+$). The
relative composition of the small particles is  $x_{\rm B} = 0.4$.
For a more detailed explanation of the parameters, see \cite{assoud2009}.
From Ref.\ \cite{assoud2009}.}
\label{fig:3}
\end{figure}

\subsection{Crystallization at system boundaries}

Similar quenches for magnetic mixtures were studied by Brownian dynamics simulations near a structured wall 
\cite{Assoud_2011} which is modeled by fixed particles on an alternating binary
equimolar square lattice cut along the (10) direction. This wall
favors local crystallites which pick up the square symmetry of the substrate. The equilibrium
state is an alternating square lattice which coincides exactly with that imposed by the external wall. After the quench,
 it is found that the number
and structure of crystallites near the walls strongly depend on the wall pattern. Even though local square
structures are favored energetically and the equilibrium state is an alternating square lattice, the number of
triangular crystallites close to the wall which has outermost fixed small particles is significantly higher than in the
unconfined case.  This effect is not
contained in classical nucleation theory.


\section{Two-dimensional colloidal mixtures in magnetic fields and under gravity}
\label{sec:2dmix}
Exposing the binary magnetic mixture described in sec.~\ref{sec:crystallization} to an in-plane homogeneous gravitational force perpendicular to the magnetic field $B$ leads to interesting newly emerging phenomena. Experimentally, this gravitational force can be realized by tilting the hanging droplet. The two components A, B of the mixture differ in both mass $M$ and magnetic susceptibility $\chi$, where A is chosen as the
heavier and more strongly coupled species. Thus, the system is characterized by the dipolar ratio $\tilde m = m_{\rm{B}}/m_{\rm{A}}$ and the mass ratio $\tilde M = M_{\rm{B}}/M_{\rm{A}}$ with $0 \leq \tilde m, \tilde M \leq 1$. First, the case of a static gravitational field was studied by Monte Carlo computer simulations and mean--field density functional theory. Second, the binary magnetic mixture was examined in the nonequilibrium situation of oscillatory gravity, which is a simple model of colloidal shaking. Thereby, Brownian dynamics simulations and
dynamic density functional theory were used to study the dynamic response of the system \cite{KruppaJCP}. 


\subsection{Static gravity: colloidal brazil--nut effect}
Monte Carlo (MC) simulations were performed for various mass ratios $0 \leq \tilde M \leq 1 $, while
the dipolar ratio was fixed to $\tilde m = 0.1$ according to recent experimental samples \cite{assoud2009,Ebert_EPJE_2008}. Thereby, a very distinct behavior of the sedimenting mixture could be observed. While for very asymmetric masses, the lighter B-particles are on top of the heavier A-particles as expected, the behavior is reversed for intermediate B-particle masses: here, the heavier A-particles are on top of the lighter B--particles. At first glance, this opposite trend is counterintuitive. In some analogy to granulate matter, it is called the (colloidal) \textit{brazil--nut effect} \cite{Esztermann2004_EPL}.

The mechanism behind this effect can be explained using an intuitive picture:
due to their strong repulsive interaction, particles of species A create a depletion zone of less repulsive particles around them
reminiscent of a bubble. Applying Archimedes' principle effectively to this bubble, an A-particle can be lifted in a fluid background of B-particles.
This ``depletion bubble'' mechanism results in the brazil--nut effect, where the heavier A-particles float on top of the lighter
B-particles. 

\begin{figure}
\centering \resizebox{0.75\columnwidth}{!}{
\includegraphics{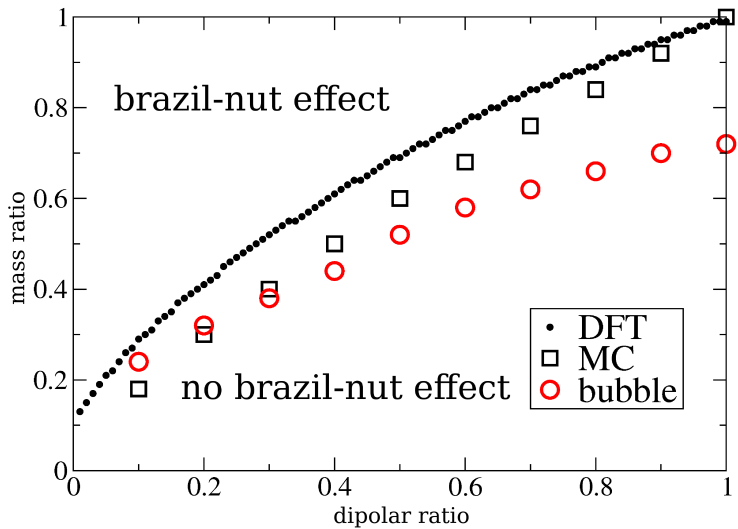} }
\caption{Separation line between the occurrence of the
colloidal brazil--nut effect and the absence of this effect in the
parameter space of dipolar ratio $\tilde m$ and mass ratio $\tilde M$. Monte
Carlo simulation data (contoured white squares), density functional data
(full black circles) and the transition line implied by the depletion bubble
picture (contoured circles) are shown. For other parameters, see Ref. \cite{KruppaJCP}. From Ref. \cite{KruppaJCP}.}
\label{fig:4}
\end{figure}

By systematically scanning the entire parameter space $0 \leq \tilde m \leq 1 $,  $0 \leq \tilde M \leq 1 $, the
line separating the brazil--nut effect from the ordinary behavior (no brazil--nut effect) was mapped, see Fig.~\ref{fig:4}. Density functional theory predictions are in good agreement with the MC simulation results, predicting the same
trends and the same slope of the separation line in the parameter space of mass and dipolar ratio.
Additionally, the intuitive depletion bubble picture provides a simple theory for the separation line based on an effective buoyancy criterion, which is included in Fig.~\ref{fig:4} and reproduces the simulation data pretty well.\\

\begin{figure}
\centering \resizebox{0.5\columnwidth}{!}{
\includegraphics{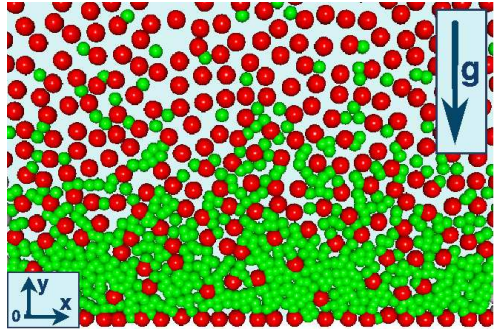} }
\caption{Simulation snapshot for $\tilde m=0.1$, $\tilde M=0.5$ showing
the marked-off bottom layer of heavy A-particles (large red spheres) at
$y = 0$ beneath the fluid of light B-particles (small green spheres).
The arrow indicates the direction of gravity, $-y$. From Ref. \cite{KruppaJCP}.}
\label{fig:5}
\end{figure}

\begin{figure}
\centering \resizebox{0.5\columnwidth}{!}{
\includegraphics{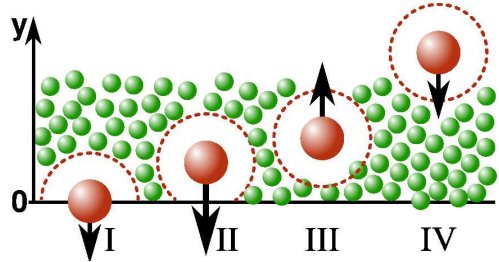} }
\caption{Schematic illustration of the metastable trapping
of A-particles at $y = 0$, leading to the formation of a boundary
layer. A-particles are represented by large red spheres, while
B-particles are depicted as smaller green spheres. Note that particles are represented as spheres with a finite radius for clarity only while in the computer simulation, point particles are considered. The solid line indicates $y = 0$, while the orientation is analogous to Fig.~\ref{fig:5}. The depletion zone surrounding A-particles is
indicated by a dashed outline.  From Ref. \cite{KruppaJCP}.}
\label{fig:6}
\end{figure}

The depletion bubble picture also implies a layering of A-particles close to the hard bottom wall of the confining container (at $y = 0$), which is
demonstrated by an actual simulation snapshot shown in Fig.~\ref{fig:5}. This effect is due to an effective attraction of an A-particle towards the hard container bottom wall:
if a single A-particle is fixed at a given distance from the bottom wall, its depletion bubble is reduced since the void space is cut by the hard
wall, see the sketch in Fig.~\ref{fig:6}. Since the A-particle is point-like, it can approach the wall very closely. Note that in Fig. \ref{fig:6}, particles are represented as spheres with finite radii for clarity, and the center of the sphere corresponds to the position of the point particle. In the computer simulation, the hard wall is implemented by systematically rejecting particle moves beyond the wall, such that the position of the point particle (i.e., the center of the spheres in Fig. \ref{fig:6}) is restricted to $y > 0$. Experimentally, this wall could be realized by a lithographic wall on a substrate or by an array of tweezers. If the A-particle is close to the wall,
the void space is half of  the full circle in the bulk (situations I and III in Fig.~\ref{fig:6}). If the height $y$ of the A-particles
increases, the depletion bubble area grows, which causes two opposing effects: first, in order to increase the depletion bubble
area, work against the osmotic pressure of the fluid B-particles is necessary, which leads to an effective attraction of the particle to the wall. 
Second, the effective buoyancy of the bubble containing the A-particle leads to a repulsive force with respect to the wall.

\subsection{Time-dependent gravity}
Furthermore, the binary magnetic mixture was examined under
time-dependent gravity, where the gravitational potential was conveniently modeled as a stepwise constant function of time. In particular, the case of a non-zero time-average of the gravity was considered.
Brownian dynamics (BD) simulations and dynamic density functional theory (DDFT) were used to study the dynamic response of the system. In an experimental setup, the case of a time-dependant in-plane gravitational force could be realized by periodically tilting the hanging droplet in opposite directions.

\begin{figure}
\centering \resizebox{0.75\columnwidth}{!}{
\includegraphics{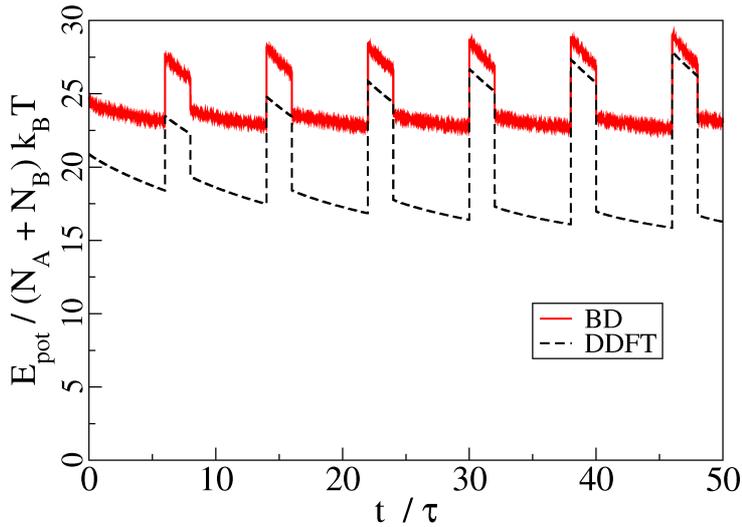} }
\caption{Total potential energy $E_{\text{pot}}/(N_{\rm{A}} + N_{\rm{B}})k_BT$ per particle
versus reduced time $t/\tau$. The parameters are $\tilde m= 0.1, \tilde M= 0.24$. From Ref. \cite{KruppaJCP}.}
\label{fig:7}
\end{figure}

The relaxation of an initially homogeneous (but interacting) fluid of A- and B-particles
towards its periodic steady  state can be monitored by observing the
instantaneous ensemble-averaged total potential energy $E_{\text{pot}}$ of the
system \cite{assoud2009}. This quantity is shown in Fig.~\ref{fig:7}, indicating that only few oscillations are needed to
get into the steady behavior. Due to the homogeneous starting configuration, the energy oscillation amplitude increases with time.
DDFT describes all trends correctly and also provides reasonable data for the potential energies and the associated
relaxation time.

\begin{figure}
\centering \resizebox{0.75\columnwidth}{!}{
\includegraphics{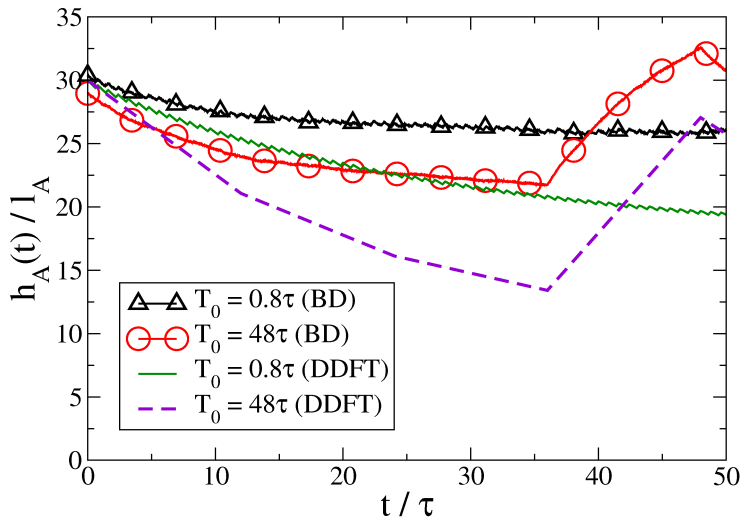} }
\caption{Time-evolution of the reduced mean A-particle height $h_{\rm{A}}(t)$ versus
reduced time $t/\tau$ for different shaking periods $T_0$. The parameters are $\tilde m= 0.1, \tilde M= 0.24$. From Ref. \cite{KruppaJCP}.}
\label{fig:8}
\end{figure}

The dynamical response of the whole system can be probed by examining the time-dependent averaged height of each particle species as shown for species A in Fig.~\ref{fig:8}. 
Upon shaking, the boundary layer of the A-particles persists. Comparing the computer simulation results to the density
profiles predicted by DDFT in more detail, the persistence of the boundary layer is contained by both methods.

\section{Quenched disorder: mixtures of mobile and immobile magnetic colloids}
\label{sec:quencheddisorder}
Recently, the influence of quenched disorder on the two-dimensional freezing
behavior was studied by using both video-microscopy of superparamagnetic confined colloidal particles
and computer simulations of two-dimensional repulsive
parallel dipoles \cite{Deutschlaender13}. A fraction of the particles was pinned to a substrate
providing quenched disorder. Similar to the set-up depicted in sec.~\ref{sec:intro}, the interaction strength was controlled by an external magnetic field, giving rise to
parallel dipolar interactions. In the pure case of a one component system,  the Kosterlitz--Thouless--Halperin--Nelson--Young (KTHNY) \cite{Halperin78,Young79} scenario was unambiguously confirmed \cite{Keim2007,Maret_PRL_2000}.
It predicts a two-stage melting
scenario with an intervening hexatic phase which is separated from the fluid and solid phases by two continuous transitions \cite{Strandburg88}, where
the melting process is mediated by the unbinding of thermally activated topological defects. 
In particular, the emergence of the hexatic phase is related to the dissociation of dislocation pairs into isolated dislocations \cite{Kosterlitz73}. These break translational symmetry, leading to a vanishing shear modulus. However, the \textit{orientational} symmetry remains quasi-long-range and the modulus of rotational stiffness, Frank's constant $K_A$, attains a nonvanishing value.

\begin{figure}
\centering \resizebox{0.75\columnwidth}{!}{
\includegraphics{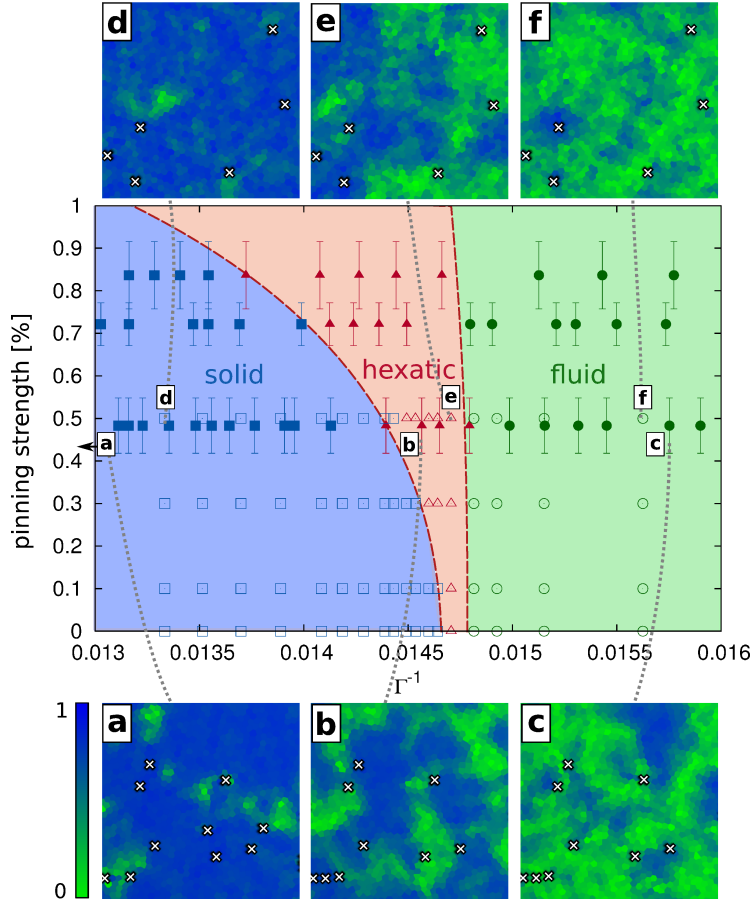} }
\caption{Phase diagram indicating solid (blue), hexatic (red), and fluid (green) phase in the parameter space of temperature $\propto 1/\Gamma$ and pinning strength, which corresponds to the percentage of immobile particles. Full symbols represent experimental data, while contoured points correspond to simulation results. The snapshots show the orientational order in the experimental (a-c) and simulated system (d-f). Only a small section of the system is shown. Voronoi cells are color-coded based on the value of the time-averaged orientational order parameter $\left\langle\left|\psi_{6}\right|\right\rangle_{t}$ according to the color bar on the left. Crosses indicate the position of immobile particles. From Ref. \cite{Deutschlaender13}.}
\label{fig:9}
\end{figure}

By systematically increasing the fraction of pinned particles, the freezing process was studied in the presence of disorder.
The occurrence of the KTHNY scenario with an intermediate hexatic phase was confirmed even for a system with quenched disorder. 
The hexatic phase was detected by analyzing the temporal correlation $g_6(t)$ of the bond order parameter \cite{nelson_book1983}.
It decays exponentially with time in the fluid and algebraically in the hexatic phase, while it reaches a constant value in the solid. 
The data obtained by Monte Carlo computer simulation and experiment were mapped to the parameter 
plane of temperature and pinning strength to obtain a phase diagram (Fig.~\ref{fig:9}) in which the transition lines for the solid-hexatic 
and hexatic-fluid transition are resolved. While the fluid-hexatic transition remains largely unaffected by disorder,
the hexatic-solid transition shifts towards lower temperatures for increasing disorder
resulting in a significantly broadened stability range of the hexatic phase.

Extracting an ``effective'' $K_A$, the scaling of the elasticity modulus was recovered in the presence of disorder. Thereby, evidence was found that melting in the presence of disorder is governed by the same defect-mediated process predicted and confirmed for pure systems.\\

\section{Conclusions}
\label{sec:conclusions}

In conclusion, binary mixtures of superparamagnetic colloids at a pending air-water interface are 
excellent model systems to explore freezing and glass formation in two-dimensional systems, 
also at one-dimensional interfaces and with quenched disorder. It is evident that in the future more complex problems can be solved
including the effect of linear shear flow \cite{Auer2004} and the correct incorporation of hydrodynamic 
interactions between the colloidal particles, which affect the non-equilibrium microstructure. The latter was recently incorporated in a strictly two-dimensional system for 
nonequilibrium band formation in oscillatory-driven mixtures \cite{Wysocki_JPCM_2011}. It is important to note that hydrodynamic interactions in the quasi-two-dimensional monolayer are different from those in a bulk suspension \cite{Cichocki2004}, where during sedimentation, the displaced fluid can flow back `above' the particle monolayer.
Also, the long-time dynamics of fluid demixing in
nonadditive systems would be an interesting system, which has also been explored in a complex plasma 
\cite{Wysocki10}. 

Tilting the plane may lead to brazil--nut effects in equilibrium.
But a time-dependent tilt can also result in transient laning of the two particle species. Lane formation was 
first found in computer simulations \cite{Dzubiella02,lowen2010instabilities} and more recent simulations in two dimensions have shown that 
laning is a continuous transition \cite{GlanzJPCM2012}. Moreover, a microscopic theory 
for laning was constructed based on the Smoluchowski equation \cite{KohlJPCM2012}.

Magnetic particles can be confined to cavities of various shapes \cite{BubeckEPL,Magnot_PCCP_2004}. It would be interesting to study the 
freezing and glass transition in finite systems contained in cavities \cite{Nemeth1998,Nemeth1999}. 
This would provide an interesting link to complex plasmas where particle
 clusters in harmonic confinement are studied intensely \cite{ivlev2012complex}.
Another relevant topic is the transport through channels \cite{Nikoubashman2010_JCP}.

The particles can also be exposed to a periodic laser-optical field allowing for new crystalline structures in equilibrium \cite{NeuhausPRL2013}. 
Finally, magnetic particles can be made active \cite{Bibette2005}. Two-dimensional swimmers were realized in this way and recent 
theoretical predictions for the
trajectory statistics in the bulk \cite{tenHagen} were confirmed 
\cite{Zheng2013_PRE}. Furthermore, self-propelled particles in shear 
flow  \cite{tenHagenPRE2011} and active crystals \cite{Menzel,Bialke} were studied.
Recently, the theoretical predictions for circle swimmers were experimentally verified by measuring the trajectories of colloidal particles self-propelled by external light fields \cite{KuemmelPRL2013}. Moreover, kinetic clustering of these self-propelled particles was found \cite{Buttinoni2013_PRL}.



\begin{acknowledgement}
 We thank P.~Keim, G.~Maret, F.~Ebert, and S.~Deutschl\"ander for the fruitful collaboration. 
This work was supported by the DFG within SFB TR6 (project C3).
\end{acknowledgement}

\bibliography{codef_refs}

\begin{thebibliography}{10}

\bibitem{ivlev2012complex}
{\sc A.~Ivlev}, {\sc H.~L{\"o}wen}, {\sc G.~Morfill}, and {\sc C.~P. Royall},
\newblock {\em Complex Plasmas and Colloidal Dispersions: Particle-Resolved
  Studies of Classical Liquids and Solids},
\newblock Series in Soft Condensed Matter, World Scientific, 2012.

\bibitem{Lowen2001}
{\sc H.~L{\"o}wen},
\newblock {\em J. Phys.: Condens. Matter} {\bf 13}, R415 (2001).

\bibitem{Lowen2009}
{\sc H.~L{\"o}wen},
\newblock {\em J. Phys.: Condens. Matter} {\bf 21}, 474203 (2009).

\bibitem{Maret_PRL_1997}
{\sc K.~Zahn}, {\sc J.~M. Mendez-Alcaraz}, and {\sc G.~Maret},
\newblock {\em Phys. Rev. Lett.} {\bf 79}, 175 (1997).

\bibitem{Zahn1999}
{\sc K.~Zahn} and {\sc G.~Maret},
\newblock {\em Curr. Opin. Colloid Interface Sci.} {\bf 4}, 60 (1999).

\bibitem{PalbergJPCM2005}
{\sc H.~L{\"o}wen}, {\sc R.~Messina}, {\sc N.~Hoffmann}, {\sc C.~N. Likos},
  {\sc C.~Eisenmann}, {\sc P.~Keim}, {\sc U.~Gasser}, {\sc G.~Maret}, {\sc
  R.~Goldberg}, and {\sc T.~Palberg},
\newblock {\em J. Phys.: Condens. Matter} {\bf 17}, S3379 (2005).

\bibitem{EPJEST_15}
{\sc P.~Dillmann}, {\sc G.~Maret}, and {\sc P.~Keim},
\newblock {\em Eur. Phys. J. Special Topics (accepted)} .

\bibitem{Froltsov1}
{\sc V.~A. Froltsov}, {\sc R.~Blaak}, {\sc C.~N. Likos}, and {\sc
  H.~L{\"o}wen},
\newblock {\em Phys. Rev. E} {\bf 68}, 061406 (2003).

\bibitem{Maret_PRL_1999}
{\sc K.~Zahn}, {\sc R.~Lenke}, and {\sc G.~Maret},
\newblock {\em Phys. Rev. Lett.} {\bf 82}, 2721 (1999).

\bibitem{Gruenberg2004}
{\sc H.~H. von Gr{\"u}nberg}, {\sc P.~Keim}, {\sc K.~Zahn}, and {\sc G.~Maret},
\newblock {\em Phys. Rev. Lett.} {\bf 93}, 255703 (2004).

\bibitem{Keim2007}
{\sc P.~Keim}, {\sc G.~Maret}, and {\sc H.~H. von Gr\"unberg},
\newblock {\em Phys. Rev. E} {\bf 75}, 031402 (2007).

\bibitem{Eisenmann2005}
{\sc C.~Eisenmann}, {\sc U.~Gasser}, {\sc P.~Keim}, {\sc G.~Maret}, and {\sc
  H.~H. von Gr\"unberg},
\newblock {\em Phys. Rev. Lett.} {\bf 95}, 185502 (2005).

\bibitem{Keim04}
{\sc P.~Keim}, {\sc G.~Maret}, {\sc U.~Herz}, and {\sc H.~H. {von
  Gr{\"u}nberg}},
\newblock {\em Phys. Rev. Lett.} {\bf {92}}, 215504 ({2004}).

\bibitem{Eisenmann2004}
{\sc C.~Eisenmann}, {\sc P.~Keim}, {\sc U.~Gasser}, and {\sc G.~Maret},
\newblock {\em J. Phys.: Condens. Matter} {\bf 16}, S4095 (2004).

\bibitem{Loewen_1994}
{\sc H.~L{\"o}wen},
\newblock {\em Phys. Rep.} {\bf 237}, 249 (1994).

\bibitem{seed}
{\sc S.~{van Teeffelen}}, {\sc C.~N. Likos}, and {\sc H.~L{\"o}wen},
\newblock {\em Phys. Rev. Lett.} {\bf 100}, 108302 (2008).

\bibitem{Sven}
{\sc S.~{van Teeffelen}}, {\sc R.~Backofen}, {\sc A.~Voigt}, and {\sc
  H.~L{\"o}wen},
\newblock {\em Phys. Rev. E} {\bf 79}, 051404 (2009).

\bibitem{Froltsov2}
{\sc V.~A. Froltsov}, {\sc C.~N. Likos}, {\sc H.~L{\"o}wen}, {\sc
  C.~Eisenmann}, {\sc U.~Gasser}, {\sc P.~Keim}, and {\sc G.~Maret},
\newblock {\em Phys. Rev. E} {\bf 71}, 031404 (2005).

\bibitem{Eisenmann_Maret}
{\sc C.~Eisenmann}, {\sc U.~Gasser}, {\sc P.~Keim}, and {\sc G.~Maret},
\newblock {\em Phys. Rev. Lett.} {\bf 93}, 105702 (2004).

\bibitem{konig2005}
{\sc H.~K{\"o}nig}, {\sc R.~Hund}, {\sc K.~Zahn}, and {\sc G.~Maret},
\newblock {\em Eur. Phys. J. E} {\bf 18}, 287 (2005).

\bibitem{Naegele_EPL_2002}
{\sc M.~Kollmann}, {\sc R.~Hund}, {\sc B.~Rinn}, {\sc G.~N{\"a}gele}, {\sc
  K.~Zahn}, {\sc H.~K{\"o}nig}, {\sc G.~Maret}, {\sc R.~Klein}, and {\sc
  J.~K.~G. Dhont},
\newblock {\em Europhys. Lett} {\bf 58}, 919 (2002).

\bibitem{Hoffmann_PRL_2006}
{\sc N.~Hoffmann}, {\sc F.~Ebert}, {\sc C.~N. Likos}, {\sc H.~L{\"o}wen}, and
  {\sc G.~Maret},
\newblock {\em Phys. Rev. Lett.} {\bf 97}, 078301 (2006).

\bibitem{Hoffmann_JPCM_2006}
{\sc N.~Hoffmann}, {\sc C.~N. Likos}, and {\sc H.~L{\"o}wen},
\newblock {\em J. Phys.: Condens. Matter} {\bf 18}, 10193 (2006).

\bibitem{hansen}
{\sc J.-P. Hansen} and {\sc I.~R. MacDonald},
\newblock {\em Theory of Simple Liquids},
\newblock Academic, London, 2006.

\bibitem{Lowen2012_Advances}
{\sc H.~L{\"o}wen}, {\sc E.~C. O\u{g}uz}, {\sc L.~Assoud}, and {\sc
  R.~Messina},
\newblock {\em Advances in Chemical Physics} {\bf 148}, 225 (2012).

\bibitem{Assoud_JPCM}
{\sc L.~Assoud}, {\sc F.~Ebert}, {\sc P.~Keim}, {\sc R.~Messina}, {\sc
  G.~Maret}, and {\sc H.~L{\"o}wen},
\newblock {\em J. Phys.: Condens. Matter} {\bf 21}, 464114 (2009).

\bibitem{assoud2009}
{\sc L.~Assoud}, {\sc F.~Ebert}, {\sc P.~Keim}, {\sc R.~Messina}, {\sc
  G.~Maret}, and {\sc H.~L{\"o}wen},
\newblock {\em Phys. Rev. Lett.} {\bf 102}, 238301 (2009).

\bibitem{Ebert_EPJE_2008}
{\sc F.~Ebert}, {\sc P.~Keim}, and {\sc G.~Maret},
\newblock {\em Eur. Phys. J. E} {\bf 26}, 161 (2008).

\bibitem{Koenig_EPL_2005}
{\sc H.~K{\"o}nig},
\newblock {\em Europhys. Lett.} {\bf 71}, 838 (2005).

\bibitem{Assoud_2007}
{\sc L.~Assoud}, {\sc R.~Messina}, and {\sc H.~L{\"o}wen},
\newblock {\em Europhys. Lett.} {\bf 80}, 48001 (2007).

\bibitem{Fornleitner_SM_2008}
{\sc J.~Fornleitner}, {\sc F.~{F. Lo Verso}}, {\sc G.~Kahl}, and {\sc C.~N.
  Likos},
\newblock {\em Soft Matter} {\bf 4}, 480 (2008).

\bibitem{Fornleitner2009_Langmuir}
{\sc J.~Fornleitner}, {\sc F.~Lo~Verso}, {\sc G.~Kahl}, and {\sc C.~N. Likos},
\newblock {\em Langmuir} {\bf 25}, 7836 (2009).

\bibitem{Fornleitner2010_PRE}
{\sc J.~Fornleitner}, {\sc G.~Kahl}, and {\sc C.~N. Likos},
\newblock {\em Phys. Rev. E} {\bf 81}, 060401 (2010).

\bibitem{Chremos2009_JPhysChemB}
{\sc A.~Chremos} and {\sc C.~N. Likos},
\newblock {\em J. Phys. Chem. B} {\bf 113}, 12316 (2009).

\bibitem{Lonetti2011_PRL}
{\sc B.~Lonetti}, {\sc M.~Camargo}, {\sc J.~Stellbrink}, {\sc C.~N. Likos},
  {\sc E.~Zaccarelli}, {\sc L.~Willner}, {\sc P.~Lindner}, and {\sc
  D.~Richter},
\newblock {\em Phys. Rev. Lett.} {\bf 106}, 228301 (2011).

\bibitem{Dillmann2008}
{\sc P.~Dillmann}, {\sc G.~Maret}, and {\sc P.~Keim},
\newblock {\em J. Phys.: Condens. Matter} {\bf 20}, 404216 (2008).

\bibitem{chowdhury1985laser}
{\sc A.~Chowdhury}, {\sc B.~J. Ackerson}, and {\sc N.~A. Clark},
\newblock {\em Phys. Rev. Lett.} {\bf 55}, 833 (1985).

\bibitem{Sperl_PRE_2007}
{\sc M.~Bayer}, {\sc J.~M. Brader}, {\sc F.~Ebert}, {\sc M.~Fuchs}, {\sc
  E.~Lange}, {\sc G.~Maret}, {\sc R.~Schilling}, {\sc M.~Sperl}, and {\sc J.~P.
  Wittmer},
\newblock {\em Phys. Rev. E} {\bf 76}, 011508 (2007).

\bibitem{Onuki_PRE_2007}
{\sc T.~Hamanaka} and {\sc A.~Onuki},
\newblock {\em Phys. Rev. E} {\bf 75}, 041503 (2007).

\bibitem{Harrowell_PRL_2006}
{\sc A.~Widmer-Cooper} and {\sc P.~Harrowell},
\newblock {\em Phys. Rev. Lett.} {\bf 96}, 185701 (2006).

\bibitem{Tanaka_PRL_2007}
{\sc T.~Kawasaki}, {\sc T.~Araki}, and {\sc H.~Tanaka},
\newblock {\em Phys. Rev. Lett.} {\bf 99}, 215701 (2007).

\bibitem{Assoud_2011}
{\sc L.~Assoud}, {\sc R.~Messina}, and {\sc H.~L{\"o}wen},
\newblock {\em Mol. Phys.} {\bf 109}, 1385 (2011).

\bibitem{KruppaJCP}
{\sc T.~Kruppa}, {\sc T.~Neuhaus}, {\sc R.~Messina}, and {\sc H.~L{\"o}wen},
\newblock {\em J. Chem. Phys.} {\bf 136}, 134106 (2012).

\bibitem{Esztermann2004_EPL}
{\sc A.~Esztermann} and {\sc H.~L\"owen},
\newblock {\em Europhys. Lett.} {\bf 68}, 120 (2004).

\bibitem{Deutschlaender13}
{\sc S.~Deutschl\"ander}, {\sc T.~Horn}, {\sc H.~L\"owen}, {\sc G.~Maret}, and
  {\sc P.~Keim},
\newblock {\em Phys. Rev. Lett.} {\bf 111}, 098301 (2013).

\bibitem{Halperin78}
{\sc B.~Halperin} and {\sc D.~R. Nelson},
\newblock {\em Phys. Rev. Lett.} {\bf 41}, 121 (1978).

\bibitem{Young79}
{\sc A.~P. Young},
\newblock {\em Phys. Rev. B} {\bf 19}, 1855 (1979).

\bibitem{Maret_PRL_2000}
{\sc K.~Zahn} and {\sc G.~Maret},
\newblock {\em Phys. Rev. Lett.} {\bf 85}, 3656 (2000).

\bibitem{Strandburg88}
{\sc K.~J. Strandburg},
\newblock {\em Rev. Mod. Phys.} {\bf 60}, 161 (1988).

\bibitem{Kosterlitz73}
{\sc J.~M. Kosterlitz} and {\sc D.~J. Thouless},
\newblock {\em J. Phys. C} {\bf 6}, 1181 (1973).

\bibitem{nelson_book1983}
{\sc D.~R. Nelson},
\newblock {\em Phase Transition and Critical Phenomena},
\newblock Academic Press, London, 1983.

\bibitem{Auer2004}
{\sc R.~Blaak}, {\sc S.~Auer}, {\sc D.~Frenkel}, and {\sc H.~L{\"o}wen},
\newblock {\em Phys. Rev. Lett.} {\bf 93}, 068303 (2004).

\bibitem{Wysocki_JPCM_2011}
{\sc A.~Wysocki} and {\sc H.~L{\"o}wen},
\newblock {\em J. Phys.: Condens. Matter} {\bf {23}}, 284117 ({2011}).

\bibitem{Cichocki2004}
{\sc B.~Cichocki}, {\sc M.~L. Ekiel-Jezewska}, {\sc G.~N{\"a}gele}, and {\sc
  E.~Wajnryb},
\newblock {\em Europhys. Lett.} {\bf 67}, 383 (2004).

\bibitem{Wysocki10}
{\sc A.~Wysocki}, {\sc C.~Raeth}, {\sc A.~V. Ivlev}, {\sc R.~K. S{\"u}tterlin},
  {\sc H.~M. Thomas}, {\sc S.~A. Khrapak}, {\sc S.~K. Zhdanov}, {\sc V.~E.
  Fortov}, {\sc A.~M. Lipaev}, {\sc V.~I. Molotkov}, {\sc O.~F. Petrov}, {\sc
  H.~L{\"o}wen}, and {\sc G.~E. Morfill},
\newblock {\em Phys. Rev. Lett.} {\bf {105}}, 045001 ({2010}).

\bibitem{Dzubiella02}
{\sc J.~Dzubiella}, {\sc G.~P. Hoffmann}, and {\sc H.~L{\"o}wen},
\newblock {\em Phys. Rev. E} {\bf {65}}, 021402 ({2002}).

\bibitem{lowen2010instabilities}
{\sc H.~L{\"o}wen},
\newblock {\em Soft Matter} {\bf 6}, 3133 (2010).

\bibitem{GlanzJPCM2012}
{\sc T.~Glanz} and {\sc H.~L{\"o}wen},
\newblock {\em J. Phys.: Condens. Matter} {\bf 24}, 464114 (2012).

\bibitem{KohlJPCM2012}
{\sc M.~Kohl}, {\sc A.~V. Ivlev}, {\sc P.~Brandt}, {\sc G.~E. Morfill}, and
  {\sc H.~L{\"o}wen},
\newblock {\em J. Phys.: Condens. Matter} {\bf 24}, 464115 (2012).

\bibitem{BubeckEPL}
{\sc R.~Bubeck}, {\sc P.~Leiderer}, and {\sc C.~Bechinger},
\newblock {\em Europhys. Lett.} {\bf 60}, 474 (2002).

\bibitem{Magnot_PCCP_2004}
{\sc K.~Mangold}, {\sc J.~Birk}, {\sc P.~Leiderer}, and {\sc C.~Bechinger},
\newblock {\em Phys. Chem. Chem. Phys} {\bf 6}, 1623 (2004).

\bibitem{Nemeth1998}
{\sc Z.~T. Nemeth} and {\sc H.~L{\"o}wen},
\newblock {\em J. Phys.: Condens. Matter} {\bf 10}, 6189 (1998).

\bibitem{Nemeth1999}
{\sc Z.~T. Nemeth} and {\sc H.~L{\"o}wen},
\newblock {\em Phys. Rev. E} {\bf 59}, 6824 (1999).

\bibitem{Nikoubashman2010_JCP}
{\sc A.~Nikoubashman} and {\sc C.~N. Likos},
\newblock {\em J. Chem. Phys.} {\bf 133}, 074901 (2010).

\bibitem{NeuhausPRL2013}
{\sc T.~Neuhaus}, {\sc M.~Marechal}, {\sc M.~Schmiedeberg}, and {\sc
  H.~L{\"o}wen},
\newblock {\em Phys. Rev. Lett.} {\bf 110}, 118301 (2013).

\bibitem{Bibette2005}
{\sc R.~Dreyfus}, {\sc J.~Baudry}, {\sc M.~L. Roper}, {\sc M.~Fermigier}, {\sc
  H.~A. Stone}, and {\sc J.~Bibette},
\newblock {\em Nature} {\bf 437}, 862 (2005).

\bibitem{tenHagen}
{\sc B.~{ten Hagen}}, {\sc S.~{van Teeffelen}}, and {\sc H.~L{\"o}wen},
\newblock {\em J. Phys.: Condens. Matter} {\bf 23}, 194119 (2011).

\bibitem{Zheng2013_PRE}
{\sc X.~Zheng}, {\sc B.~ten Hagen}, {\sc A.~Kaiser}, {\sc M.~Wu}, {\sc H.~Cui},
  {\sc Z.~Silber-Li}, and {\sc H.~L\"owen},
\newblock {\em Phys. Rev. E} {\bf 88}, 032304 (2013).

\bibitem{tenHagenPRE2011}
{\sc B.~{ten Hagen}}, {\sc R.~Wittkowski}, and {\sc H.~L{\"o}wen},
\newblock {\em Phys. Rev. E} {\bf 84}, 031105 (2011).

\bibitem{Menzel}
{\sc A.~M. Menzel} and {\sc H.~L{\"o}wen},
\newblock {\em Phys. Rev. Lett.} {\bf 110}, 055702 (2013).

\bibitem{Bialke}
{\sc J.~Bialk{\'e}}, {\sc T.~Speck}, and {\sc H.~L{\"o}wen},
\newblock {\em Phys. Rev. Lett.} {\bf 108}, 168301 (2012).

\bibitem{KuemmelPRL2013}
{\sc F.~K{\"u}mmel}, {\sc B.~ten Hagen}, {\sc R.~Wittkowski}, {\sc
  I.~Buttinoni}, {\sc R.~Eichhorn}, {\sc G.~Volpe}, {\sc H.~L{\"o}wen}, and
  {\sc C.~Bechinger},
\newblock {\em Phys. Rev. Lett.} {\bf 110}, 198302 (2013).

\bibitem{Buttinoni2013_PRL}
{\sc I.~Buttinoni}, {\sc J.~Bialk\'e}, {\sc F.~K\"ummel}, {\sc H.~L\"owen},
  {\sc C.~Bechinger}, and {\sc T.~Speck},
\newblock {\em Phys. Rev. Lett.} {\bf 110}, 238301 (2013).

\end{thebibliography}

\end{document}